\documentclass[twoside,pre,preprintnumbers,floatfix]{revtex4}

\usepackage{amsmath}
\usepackage{amssymb}
\usepackage{graphicx}
\usepackage{dcolumn}
\usepackage{bm}
\usepackage{epstopdf}

\def\epsilon{\varepsilon}
\def\theta{\vartheta}
\def\rho{\varrho}

\begin{document}

\title{Magnetic field dependence of electronic properties of MoS$_2$ quantum dots with different edges}

\author{Qiao Chen$^{1,2}$$\footnote{ Email: cqhy1127@aliyun.com}$}

\author{L. L. Li$^{2,3}$$\footnote{ Email: longlong.li@uantwerpen.be}$}
\author{F. M. Peeters$^{2}$$\footnote{ Email: francois.peeters@uantwerpen.be}$}

\affiliation{$^{1}$Department of Maths and Physics, Hunan Institute of Engineering, Xiangtan 411104, China\\
$^{2}$Department of Physics, University of Antwerp, Groenenborgerlaan 171, B-2020 Antwerp, Belgium \\
$^{3}$Key Laboratory of Materials Physics, Institute of Solid State Physics, Chinese Academy of Sciences, Hefei 230031, China}
\date{\today}
\begin{abstract}
Using the tight-binding approach, we investigate the energy spectrum of square, triangular and hexagonal MoS$_2$ quantum dots (QDs) in the presence of a perpendicular magnetic field. Novel edge states emerge in MoS$_2$ QDs, which are distributed over the whole edge which we call ring states. The ring states are robust in the presence of spin-orbit coupling (SOC). The corresponding energy levels of the ring states oscillate as function of the perpendicular magnetic field which are related to Aharonov-Bohm oscillations. Oscillations in the magnetic field dependence of the energy levels and the peaks in the magneto-optical spectrum emerge (disappear) as the ring states are formed (collapsed). The period and the amplitude of the oscillation decreases with the size of the MoS$_2$ QDs.
\end{abstract}

\maketitle

\section{Introduction}
After the initial boom in graphene research, recent years have seen a surge of interest in other two-dimensional (2D) atomic crystals\cite{Geim1}. Among them, molybdenium disulfide (MoS$_2$), a prototypical transition metal dichalcogenide (TMD), has attracted significant interest due to their extraordinary electronic and optical properties. Together with the excellent electrostatic control inherent of 2D materials, the large band gap and high carrier mobility makes them well
suited for low power electronics and optoelectronic applications\cite{ref06,ref07,ref08,ref09}. MoS$_2$ can be fabricated by, for example, mechanical exfoliation\cite{Radisavljevic}, chemical vapor deposition(CVD)\cite{Wu} or by direct growth methods\cite{Kim}.

Monolayer MoS$_2$ is in many ways similar to graphene, but it also has crucial differences, such as metal $d$ bands around the Fermi level, a direct band gap, and a lack of inversion symmetry\cite{Zhu}. Transition-metal $d$ orbitals lead to strong spin-orbit coupling (SOC) effects\cite{Xiao}. Coupled with the lack of inversion symmetry, SOC leads to a large spin splitting of the valence bands at the corners of the Brillouin zone. The splittings have opposite signs at $K$ and $K^\prime$ points, which gives rise to valley-dependent optical transitions\cite{Xiao}. This has been verified by experiment using dynamical pumping of valley polarization by circularly polarized light in monolayers of MoS$_2$\cite{Cao,Mak,Zeng}. In addition, electrons and holes have valley degrees of freedom, which may be used for information encoding and processing\cite{Xiao2,Yao1,Zhu2}. Recently, control of the valley pseudospin via an external magnetic field was demonstrated experimentally, which works by breaking the degeneracy of energy states at $\pm K$ valleys\cite{MacNeill,Aivazian}. These results suggest that monolayer MoS$_2$ can be used as a possible host for integrated spintronics and valleytronics.

Quantum dots (QDs) in novel low-dimensional structures, such as graphene\cite{GrapheneQD1,GrapheneQD2,GrapheneQD3,GrapheneQD4,GrapheneQD5} and black phosphorene\cite{BP1,BP2,BP3,BP4,BP5}, are actively studied and the applicability of these structures for hosting qubits has also been discussed. To date, several strategies have been developed for preparation of MoS$_2$ QDs, including lithium intercalation\cite{Small}, liquid exfoliation in organic solvents\cite{ACSNano}, hydrothermal synthesis\cite{JMCA}, electrochemical etching\cite{Chem}, electro-Fenton reaction processing\cite{Nanoscale}, and grinding exfoliation\cite{ACSAMI}. The electronic properties of the TMD QDs  can be measured using STM\cite{Nature}. However, to the best of our knowledge, only few theoretical works have been carried out in the field of MoS$_2$ quantum dots. Korm\'{a}nyos {\sl{ et al.}} considered quantum dots formed in transition metal dichalcogenides (TMDC) by electronic gating and calculated the magnetic field dependence of the bound states in the quantum dot\cite{MoQD1}. They found that all states are both valley and spin split, which suggests that these quantum dots can be used as valley-spin filters. Moreover, because of the relatively strong intrinsic SOC splitting in the conduction band, the spin and valley states in TMDC quantum dots can be used as quantum bits. The electronic structure of triangular MoS$_2$ quantum dots have been investigated by means of an effective $\bf{k\cdot p}$ two-band model\cite{MoQD2}. The QDs exhibit edge states, which are determined by the curvatures of conduction- and valence-band, localized on the edges and with energies lying in the band gap. Recently, a three-band model has been used to investigate the electronic structure of MoS$_2$ quantum dots by the tight-binding approach\cite{MoQD3}. It was shown that it is possible to form quantum dots with the same shape but having different electronic properties due to the orbital asymmetry. However, in those publications the effect of a perpendicular magnetic field was not considered.

Edge-structure engineering of 2D materials is an important technique to control their electronic properties. Likewise, the electrical properties of MoS$_2$ are strongly edge dependent, ranging from insulating to metallic for armchair- and zigzag-terminated edges, respectively\cite{MoS2}. Based on the above considerations, we design MoS$_2$ quantum dots with different edges and investigate their electronic properties in the presence of a perpendicular magnetic field. Our calculations show that: (i) edge states are found in the energy gap; (ii) the edge structure affects the edge states noticeable; and (iii) the energy levels of ring states oscillate with the magnetic flux. The magnitude and period of oscillations decrease with the size of QDs; and (iv) these oscillations become aperiodic when ring states mix with bulk states.

This paper is organized as follows. In section II, we present the TB Hamiltonian of monolayer MoS$_2$ and show how the magnetic field is taken into account. In
section III, we present and discuss the effects of edges, and the shape of the QDs on its electronic properties. The effect of a magnetic field is discussed in section IV. Finally, we conclude with a summary in section V.

\section{Model and theory}
\begin{figure}
\centering
\includegraphics[width=0.80\textwidth]{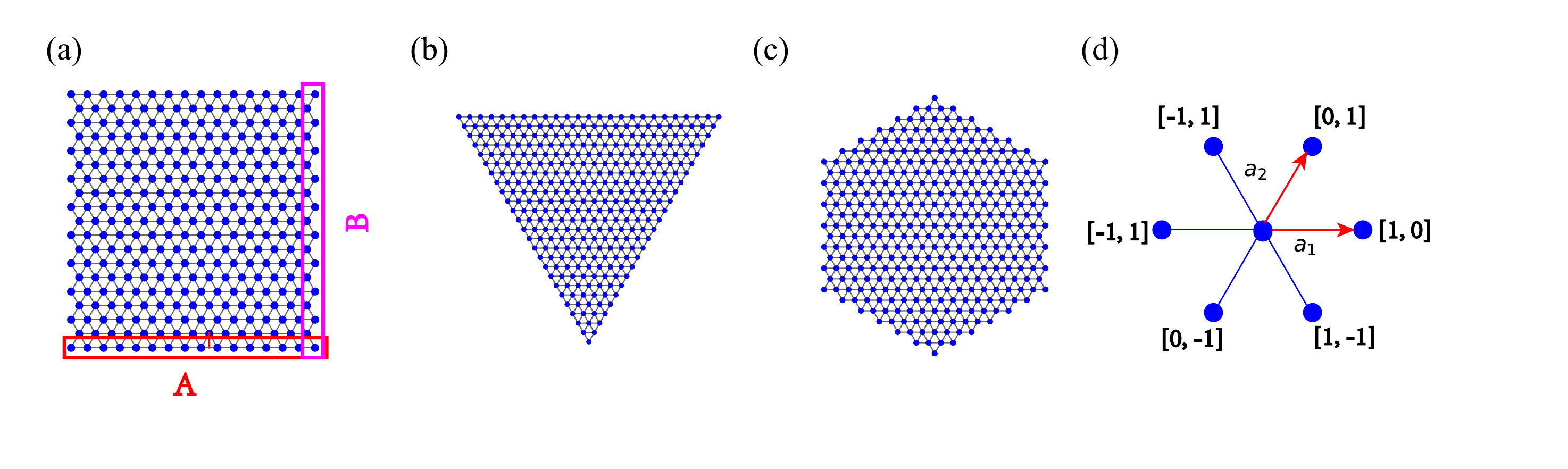}
\caption{  Sketch of the considered different shaped MoS$_2$ flakes (only the Mo plane is shown): (a) mixed edge QD, where A(B) stands for zigzag (armchair) edge, (b) triangular QD with zigzag edges, (c) hexagonal QD with armchair edges, and (d) the relative position of neighbor atoms, where $a_1$, $a_2$ are the lattice vectors. }\label{fig:fi1}
\end{figure}
Fig. 1 shows a collection of different shaped flakes with different types of edges, including three different shaped monolayer MoS$_2$ quantum dots with area
approximately equal to $27 nm^2$. The basic information including the number of edge atoms $N_{edge}$, the number of total atoms in QDs $N_{total}$, the side length $L_{edge}$ and the surface area $S$ are given in Table I for the investigated QDs.

\begin{table}[h]%
\begin{tabular*}
{0.6\textwidth}[c]{@{\extracolsep{\fill} }cccc}\hline\hline
$\mathrm{shape}$& $\mathrm{Square} (a)$ & $\mathrm{Triangular} (b)$ & $\mathrm{Hexagonal} (c)$ \\\hline
$\mathrm{N_{total}}$ & $295$ & $325$ & $343$ \\
$\mathrm{N_{edge}}$ & $48$&$72$ & $36$ \\
$\mathrm{L_{edge}/nm}$ & $5.0$& $7.66$ & $3.32$ \\
$\mathrm{S/nm^2}$ & $25$&$25.4$ & $28.6$ \\\hline\hline
\end{tabular*}
\caption{Basic information of the numerically investigated MoS$_2$ QDs including the number of edge atoms $N_{edge}$, total number of atoms $N_{total}$, the side length $L_{edge}$ and the surface area $S$. }%
\label{tabular:1}%
\end{table}

We define type A edge structure and type B edge structure in Fig. 1(a), where A(B) stands for zigzag (armchair) edge. In type A edge, every edge atom bonds with four nearest atoms. However, in type B edge each edge atom bonds with three nearest atoms. The different bonding conditions well affect the electronic structure in quantum dots.

 We use the three-band tight-binding (TB) model\cite{Yao}, the corresponding parameters are given in Table II. This approach has been used in previous studies of monolayer equilateral triangle shaped MoS$_2$ quantum dots\cite{MoQD3} in the absence of magnetic field. The tight-binding Hamiltonian of the MoS$_2$ QDs can be written as
 \begin{eqnarray}
 H=\sum_{i}\mathcal{E}_{i,\mu\nu}c_{i,\mu}^\dag c_{i,\nu}+\sum_{i\neq j}t_{ij,\mu\nu}^{mn}c_{i,\mu}^\dag c_{j,\nu}
 \end{eqnarray}
 where $i$,$j$ and $\mu$,$\nu$ run over the lattice sites and atomic orbital bases, respectively. The atomic orbital, labeled by $\mu,\nu=1,2,3$, are defined by the wave vector
 \begin{eqnarray}
\psi=(d_{z^2},d_{xy},d_{x^2-y^2})
\end{eqnarray}
in the three-band model. $m$, $n$ refer to the relative positions of two neighbor atoms, as shown in Fig. 1(d). The on-site energies are given by
\begin{eqnarray}
\mathcal{E}_{i}=
\begin{bmatrix}
\epsilon_1 & 0 & 0\\
0          & \epsilon_2 &-is\lambda\\
0&is\lambda&\epsilon_2
\end{bmatrix}.
\end{eqnarray}
\{where $\lambda$ characterizes the strength of the SOC and $s=\pm1$ is the $z$ component of the spin degree of freedom. The hopping matrices between two atoms are given by
\begin{eqnarray}
t_{ij}^{10}=
\begin{bmatrix}
t_0 & -t_1 & t_2\\
t_1          & t_{11} &-t_{12}\\
t_2&t_{12}&t_{22}
\end{bmatrix},\quad
t_{ij}^{0-1}=
\begin{bmatrix}
t_0 & \frac{t_1}{2}+\frac{\sqrt{3}t_2}{2} &\frac{\sqrt{3}t_1}{2}-\frac{t_2}{2}\\
-\frac{t_1}{2}+\frac{\sqrt{3}t_2}{2}          & \frac{t_{11}}{4}+\frac{3t_{22}}{4} &\frac{\sqrt{3}(t_{11}-t_{22})}{4}-t_{12}\\
-\frac{\sqrt{3}t_1}{2}-\frac{t_2}{2}&\frac{\sqrt{3}(t_{11}-t_{22})}{4}+t_{12}&\frac{3t_{11}}{4}+\frac{t_{22}}{4}
\end{bmatrix},
\end{eqnarray}
and
\begin{eqnarray}
t_{ij}^{1-1}=
\begin{bmatrix}
t_0 & -\frac{t_1}{2}-\frac{\sqrt{3}t_2}{2} &\frac{\sqrt{3}t_1}{2}-\frac{t_2}{2}\\
\frac{t_1}{2}-\frac{\sqrt{3}t_2}{2}          & \frac{t_{11}}{4}+\frac{3t_{22}}{4} &\frac{\sqrt{3}(t_{22}-t_{11})}{4}+t_{12}\\
-\frac{\sqrt{3}t_1}{2}-\frac{t_2}{2}&\frac{\sqrt{3}(t_{22}-t_{11})}{4}-t_{12}&\frac{3t_{11}}{4}+\frac{t_{22}}{4}
\end{bmatrix}.
\end{eqnarray}

  Here we generalized previous results to different flake shapes and add a perpendicular magnetic field. When a perpendicular magnetic field $B$ is applied to the monolayer MoS$_2$ quantum dot, the hopping energies $t_{ij}$ should be replaced by
\begin{eqnarray}
t_{ij}^{mn}\rightarrow t_{ij}^{mn}exp(i\frac{2\pi e}{h}\int_{\textbf{r}_i}^{\textbf{r}_j}\textbf{A}\cdot d\textbf{l})
\end{eqnarray}
where $h$ is Planck's constant and $\textbf{A}$ is the vector potential induced by the field $\textbf{B}$. We use the Landau gauge, with the vector potential $\textbf{A}=(0,Bx,0)$. The magnetic flux threading the MoS$_2$ unit cell is defined as $\Phi=\frac{\sqrt{3}}{4}Ba^2$ in units of the flux quantum $\Phi_0=h/e$, where $a=0.319 nm$ is the lattice constant. The magnetic flux threading the QDs is defined as $\Phi^*=BS$, where $S$ is the surface area of the QD.

\begin{table}[h]%
\begin{tabular*}
{0.6\textwidth}[c]{@{\extracolsep{\fill} }cccc}\hline\hline
 $\mathrm{Symbol}$ & $\mathrm{Value}$ & $\mathrm{Symbol}$ &$\mathrm{Value}$ \\\hline
  $\varepsilon_1$ &$1.046$ & $\varepsilon_2$ & $2.104$  \\
 $t_0$ &$-0.184$ & $t_1$ & $0.401$  \\
 $t_2$ &$0.507$ & $t_{11}$ & $0.218$  \\
 $t_{12}$ &$0.338$ & $t_{22}$ & $0.057$  \\\hline\hline
\end{tabular*}
\caption{The parameters in units of eV of the three-band TB model as determined from first-principle calculation band structure of MoS$_2$ using GGA.  }%
\label{tabular:2}%
\end{table}

The energy levels and wave functions in the monolayer MoS$_2$ QD are obtained by diagonalizing the TB Hamiltonian matrix numerically. All numerical tight-binding calculations are performed using the recently developed Pybinding package\cite{Dean}. The electronic density of states (DOS) for a QD are the sum of a series of $\delta$ functions, which we numerically calculated with a Gaussian broading as
\begin{eqnarray}
DOS(E)=\frac{1}{\sqrt{2\pi\Gamma^2}}\sum_n\exp[\frac{-(E-E_n)^2}{2\Gamma^2}]
\end{eqnarray}
where $\Gamma$ is the broadening factor and $E_n$ is the energy value for the $n$th eigenstate.

The optical properties of MoS$_2$ QDs are promising for potential applications in optoelectronic devices. The optic absorption spectra are given by
\begin{eqnarray}
A(\hbar\omega)=\sum_{i,j}(E_j-E_i)|{\bf{p \cdot M_{ij}}}|^2\delta(E_i-E_j+\hbar\omega)
\end{eqnarray}
where $E_i(E_j)$ in the energy level for the initial (final) state $|i>(|j>)$ and $\bf{p}$ the polarization of incident light. The dipole matrix element for the transition from the initial state $|i>$ to the final state $|j>$ is given by $\bf{M}_{ij}=<j|\bf{r}|i>$, where light is assume to be polarized in the $(x,y)$ plane. We define an averaged optical absorption $A_{ave}(\hbar\omega)=[A_x(\hbar\omega)+A_y(\hbar\omega)]/2$, where $A_x(\hbar\omega)(A_y(\hbar\omega))$ represents the optical absorption when light is polarized along $x$ ($y$) direction, respectively. In the numerical calculation of $A_{ave}(\hbar\omega)$, we use the same Gaussian broadening as in Eq. (7) for the delta function in Eq. (8). The three-band TB model which is employed in our work is qualitatively accurate to describe the optical absorption of monolayer MoS$_2$ which was demonstrated to have a direct band gap at the $K$ and $K^\prime$ points, and can thus capture the main physics of the optical absorption of its QDs.

\section{Shape and edge dependence}

\begin{figure}
\centering
\includegraphics[width=0.8\textwidth]{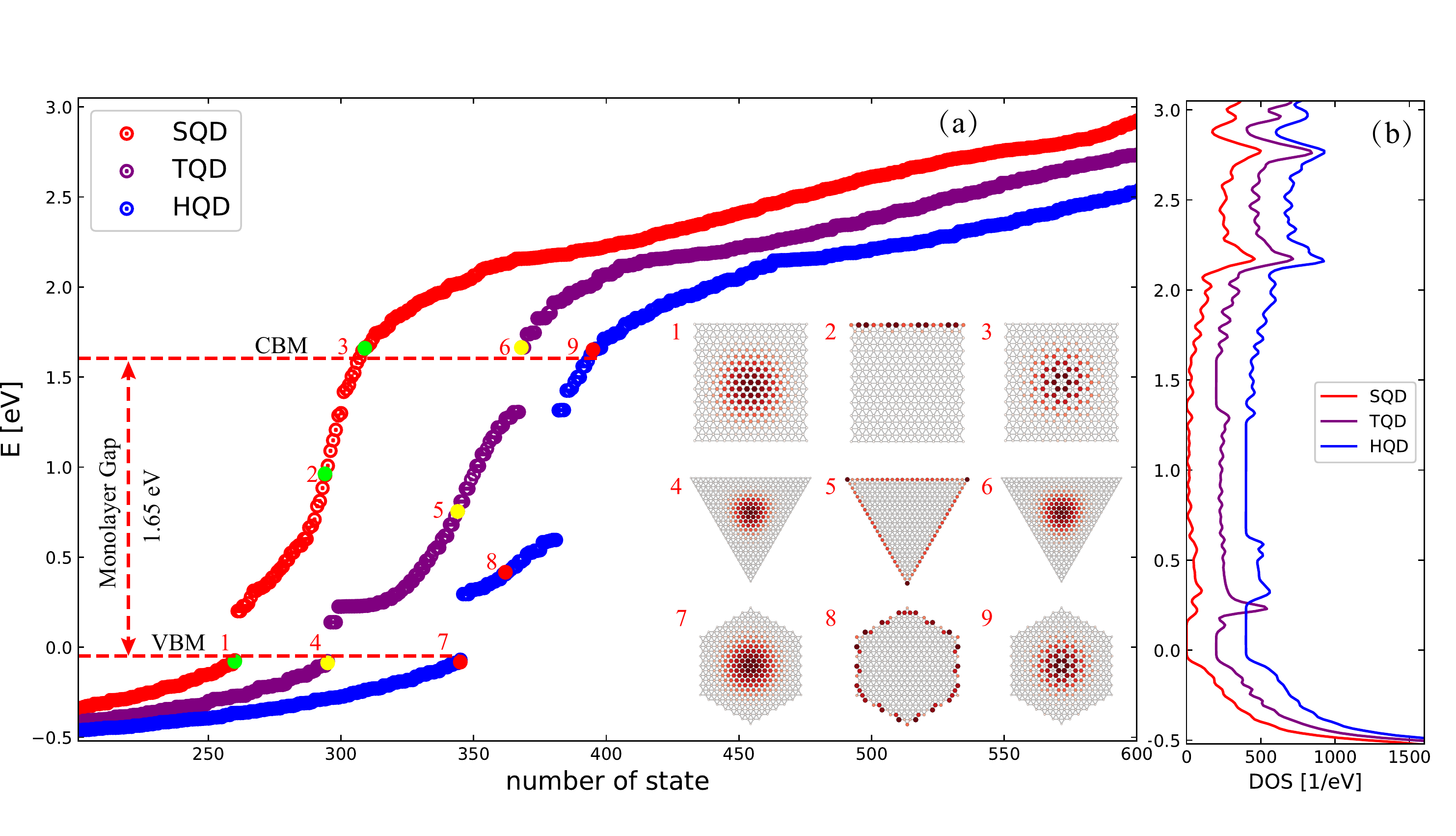}
\caption{ (a) Energy levels of quantum dots. For clarity we displaced the state index by 20 for the different curves in the horizontal direction. The probability density of the states indicated by numbers are given in the insert figures. (b) Each of the corresponding DOS of the three QDs are offset by 200 $eV^{-1}$ and we used a broadening $\Gamma=0.02eV$.  }\label{fig:fig2}
\end{figure}
For simplicity we consider the spinless model first, and the SOC is omitted, i.e., $\lambda=0$. First, we calculate the energy levels of MoS$_2$ quantum dots shown in Fig. 2 in the absence of a magnetic field. The energy gap of monolayer MoS$_2$ is $1.65 eV$ in the three-band model. Edge states emerge in the band gap, as is apparent from the probability density shown in the insets of Fig. 2. In phosphorene quantum dots, the edge states are well separated from the bulk conduction and valence bands states\cite{BP2}. A clear gap exists between valence band and edge states in MoS$_2$ QDs. However, the edge states in the gap of SQD and HQD are connected with the conduction band, i.e., bulk states. The edge states in TQD are well separated from the bulk states. Thus whether or not the edge states are separated from the conduction band depends on the shape of QDs. Although, the area of three QDs are almost the same, the number of atoms are sightly different. Therefore, the number of states is different. QDs with different shapes show different electronic structure. On the other hand, comparing the DOS of the three QDs, we see that the DOS are quite similar in the valence band, while the difference is more noticeable in the band gap where edge states are found.

In order to investigate the influence of the type of edge, we consider hexagonal quantum dots (HQDs) with different edges in the absence of magnetic field.
As can be seen from Fig. 3, the number of gap states in HQD with zigzag edge is larger than the HQD with armchair edge. The number of edge states is related with the number of edge atoms\cite{BP1,BP2}. The number of edge atoms in a quantum dot with zigzag edges is larger than for a quantum dot with armchair edges (see Table I). On the other hand, comparing the eigenenergies of the two HQDs, we see that the energy states are quite similar in the valence band and there is a tiny difference in the conduction band. However, the largest difference lies in the band gap regime where edge states dominate. Similar conclusions can be obtained by examining the density of states in Fig. 3(b). Here, we introduced a Gaussian broadening of $0.02eV$ for each energy level. The DOS of two HQDs are almost identical in the valence band, while huge difference emerges in the energy gap. We can understand these from the inset figures in Fig. 3(a). The structures in the two red circles are almost the same where bulk states exist. The states in the valence band are bulk states. Therefore, the states in the valence band are almost the same. However, the edge structure (outside of the red circles) are different. Therefore, the edge states show huge differences. In the work by Rostami et al. on nanoribbons\cite{Rostami}, metallic and gapped edge modes were obtained in the zigzag and armchair ribbons and they were explained in terms of topology using a simple two-band $k\cdot p$  model around the $K$-points. There is a topological invariant $(Z_2)$ characterizing such topology: $Z_2=1(0)$ indicates topological (conventional) states. Concerning edge states in QDs, one may find a similar topology-related explanation (nonzero Chern number per valley) in Ref. \cite{MoQD2}.  However, edge states obtained in our work are due to the boundary condition (zigzag or armchair) that is naturally incorporated in the tight-binding (TB) model. Different boundary conditions implement different edge effects in the TB model, which play an important role in nanostructures such as QDs due to their large edge/volume ratio. Moreover, edge states obtained in our work are found to be present at both armchair and zigzag boundaries (see Fig. 3(c)). However, the edge states can only distribute over the zigzag edges in graphene and phosphorene QDs\cite{GrapheneQD2,BP1,BP2}.

\begin{figure}
\centering
\includegraphics[width=0.80\textwidth]{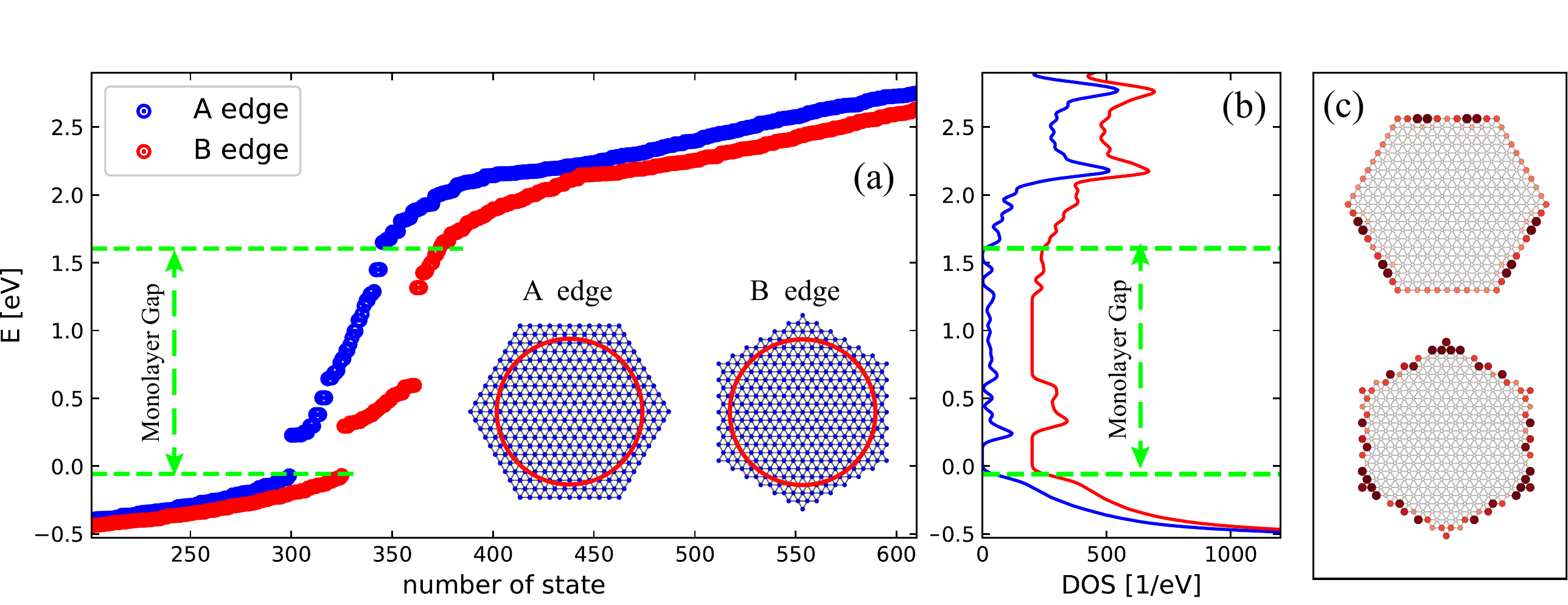}
\caption{ Hexagonal quantum dots (HQDs) with zigzag edges and armchair edges. (a) The corresponding eigenenergies and (b) DOS of the HQDs with different edges in the absence of magnetic field. (c) The probability densities of state $334$ of HQD with zigzag edge and state $325$ of HQD with armchair edge. The side length of the HQDs are about $3.32nm$. For clarity the index of state and the DOS are offset by 20 and 50 $eV^{-1}$, respectively.  }\label{fig:fig3}
\end{figure}

\section{Magnetic field dependence}

\begin{figure}
\centering
\includegraphics[width=0.85\textwidth]{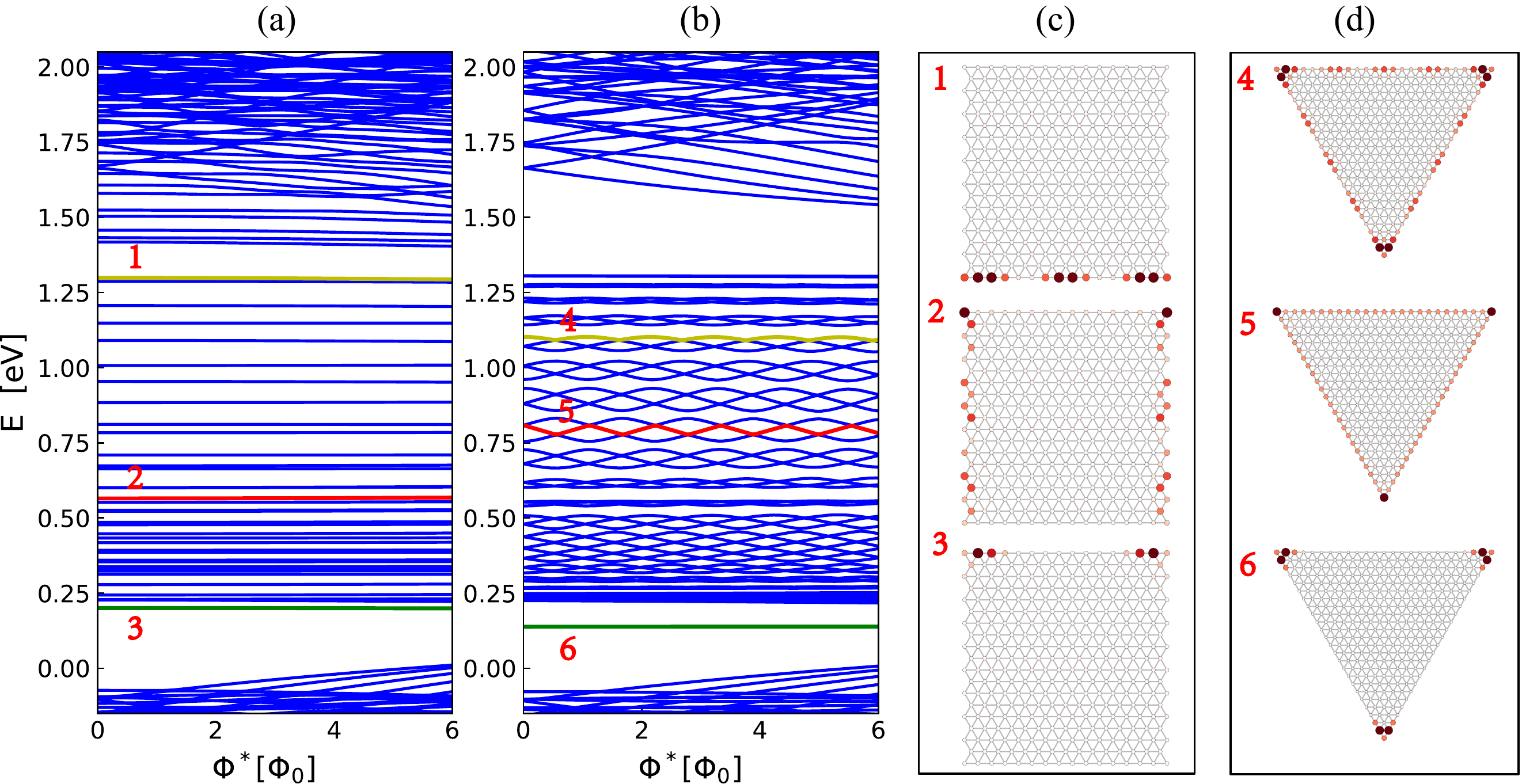}
\caption{Energy spectrum of (a) SQD and (b) TQD in the presence of magnetic field. (c) and (d) The corresponding probability densities of states marked in (a) and (b).}\label{fig:fig4}
\end{figure}
First, we investigate the energy spectrum of a SQD (Fig. 1(a)) and a TQD (Fig. 1(b)) in the presence of a perpendicular magnetic field.
The energy spectrum of gap states show the greatest differences in Figs. 4(a) and (b). In MoS$_2$ SQD, a nearly flat band is formed by the gap states in the presence of a magnetic field and with increasing magnetic field flux. However, most of the gap states in TQD show threefold oscillating behavior in the presence of the magnetic field. Previously it was shown that the energy levels in graphene QDs and phosphorene QDs do not oscillate with the magnetic flux\cite{GrapheneQD2,BP2,BP1}. It indicates that the electronic structure of MoS$_2$ is different from graphene and phosphorene. The oscillating behavior of energy levels was also found in graphene and phosphorene quantum rings, which is caused by the AB effect\cite{Peeters2,LLli}. Besides the oscillating energy levels, there are also some energy levels in TQD that are not affected by the magnetic field, such as the one shown by the green horizon line in Fig. 4(b) (see the line indicated by the number '6'). In order to explain this oscillating behavior in TQD, we give the probability densities of three arbitrary energy levels (green, red and yellow lines in Figs. 4(a) and (b)), which are marked by $1,2\cdots 6$, in Figs. 4(c) and (d).  As can be seen in Fig. 4(c), the gap states in SQD are all localized edge states. These localized edge states can be distributed at both the zigzag or armchair edge and are not affected by the magnetic field. However, the probability densities of some gap states in TQD distribute over the whole edge, and form rings as shown in Fig. 4(d). We call these states that are distributed over the whole edge as ring states. As magnetic flux thread through such a ring state, oscillating behavior emerges, like the AB oscillation caused by the enclosed magnetic flux. The state corresponding to the green horizon line in Fig. 4(b) is a localized state as show in Fig. 4(d). Therefore, it does not show any oscillating behavior. We also note that in a very recent paper\cite{Segarra}, it was found that there are a series of iso-spaced states which show a linear dependence on the magnetic field, and these states are associated with the topological character of the two-band $k\cdot p$ Hamiltonian. In order to obtain these states in Ref.\cite{Segarra}, the infinite mass boundary condition was imposed at the border of QDs. Our tight-binding results show that edge states of QDs, which arise due to the boundary condition, are present at both zigzag and armchair edges. The discrepancy between the results obtained by their two-band $k\cdot p$ model and our tight-binding model are due to: (i) The models themselves are different: the two-band $k\cdot p$ model can be viewed as an expansion of the tight-binding model up to second order in $k$; (ii) The infinite-mass boundary condition is rather artificial and is physically different from the zigzag or armchair one. The latter can model the atomic structure of QD edges whereas the former cannot. In $k\cdot p$ model approaches, the infinite-mass boundary condition has been widely used because it can give analytical results, whereas in the tight-binding (TB) model, zigzag or armchair boundary condition can be naturally introduced. Therefore, different boundary conditions are the primary cause for the discrepancy between the results obtained by their two-band $k\cdot p$ model and our tight-binding model.

The spectra of an ideal metallic ring exhibits perfect periodic AB oscillations with period $\Phi_0$, whose energy spectrum can be portrayed by the simple analytic expression
\begin{eqnarray}
E_i(\Phi)&\propto& (l-\Phi/\Phi_0)^2,
\end{eqnarray}
where $l$ is the angular quantum number which takes integer values. In contrast, the period of oscillation in TQDs is sightly larger than $\Phi_0$. Note that angular quantum number $l$ is well defined in an ideal ring. However, it is not a good quantum number in TQDs. The coupling of energy states leads to changes in the period.

 \begin{figure}
\centering
\includegraphics[width=0.75\textwidth]{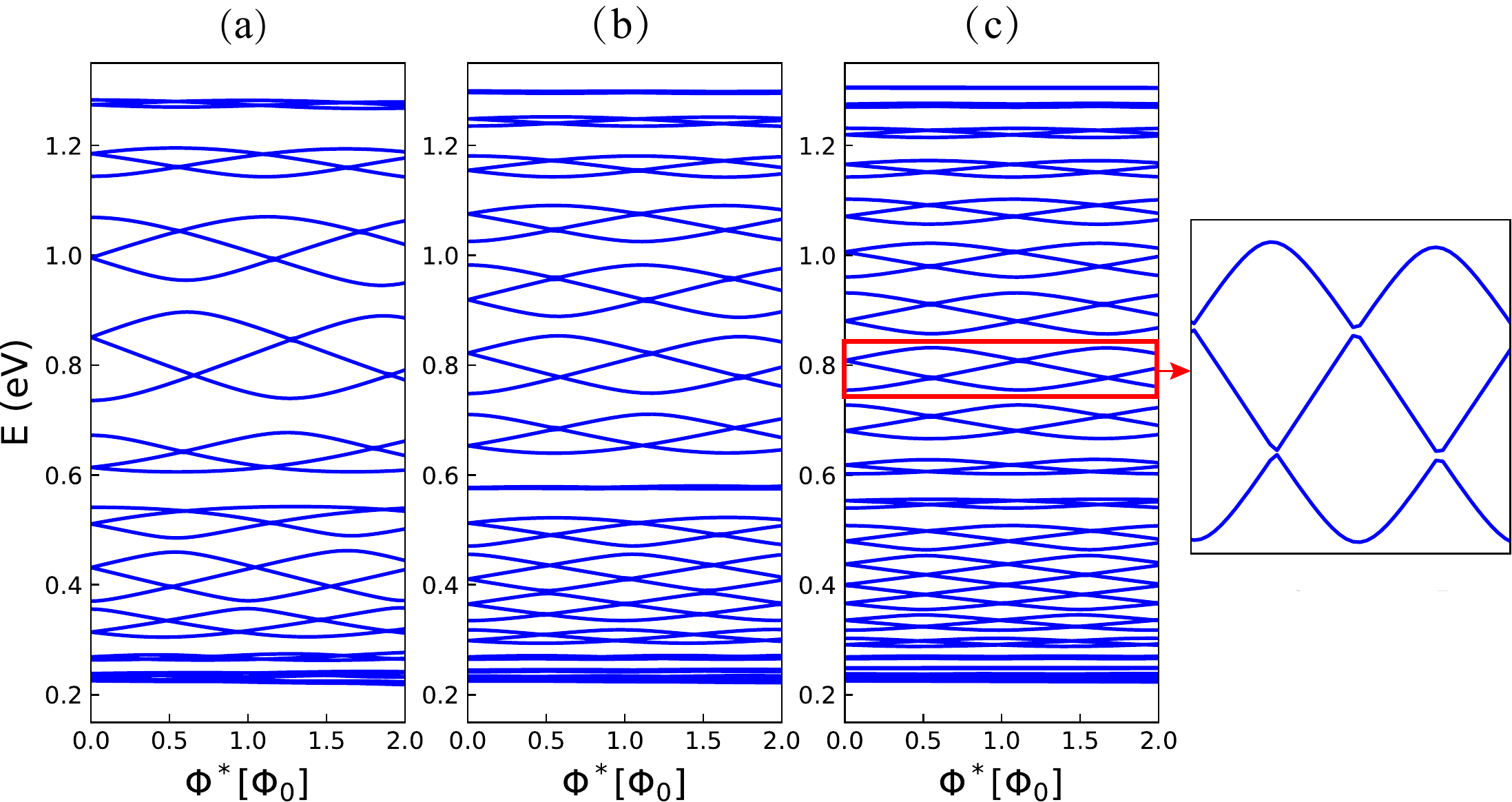}
\caption{Energy spectrum of triangular quantum dot with different size in the presence of magnetic field. (a) $L_{edge}=3.828nm$, (b) $L_{edge}=5.742nm$, (c) $L_{edge}=7.66nm$. }\label{fig:fig5}
\end{figure}

To further investigate this oscillating behavior, we investigated the energy spectrum of TQDs with different size in the presence of a perpendicular magnetic field in Fig. 5. Oscillating behavior emerge in TQDs with different size. However, the number of states that display oscillating behavior increases with the size of TQDs.  It indicates that the number of ring states increases with the size of TQDs. The reason is that the number of ring states is proportional to the number of edge atoms. The magnitude and period of the oscillation increase as the size decreases. The magnitude of oscillation is proportional to the coupling strength of ring states, i.e., $|<\Phi_1^r|\Phi_2^r>|^2$. The area which is encircled by the ring states becomes smaller as the size decreases. The coupling between ring states becomes stronger as the area decreases. Therefore, the magnitude and the period of oscillation decreases with the size of TQDs. In addition, the oscillating energy levels exhibit anticrossings as shown by the zoomed plot in Fig. 5(c).

\begin{figure}
\centering
\includegraphics[width=0.75\textwidth]{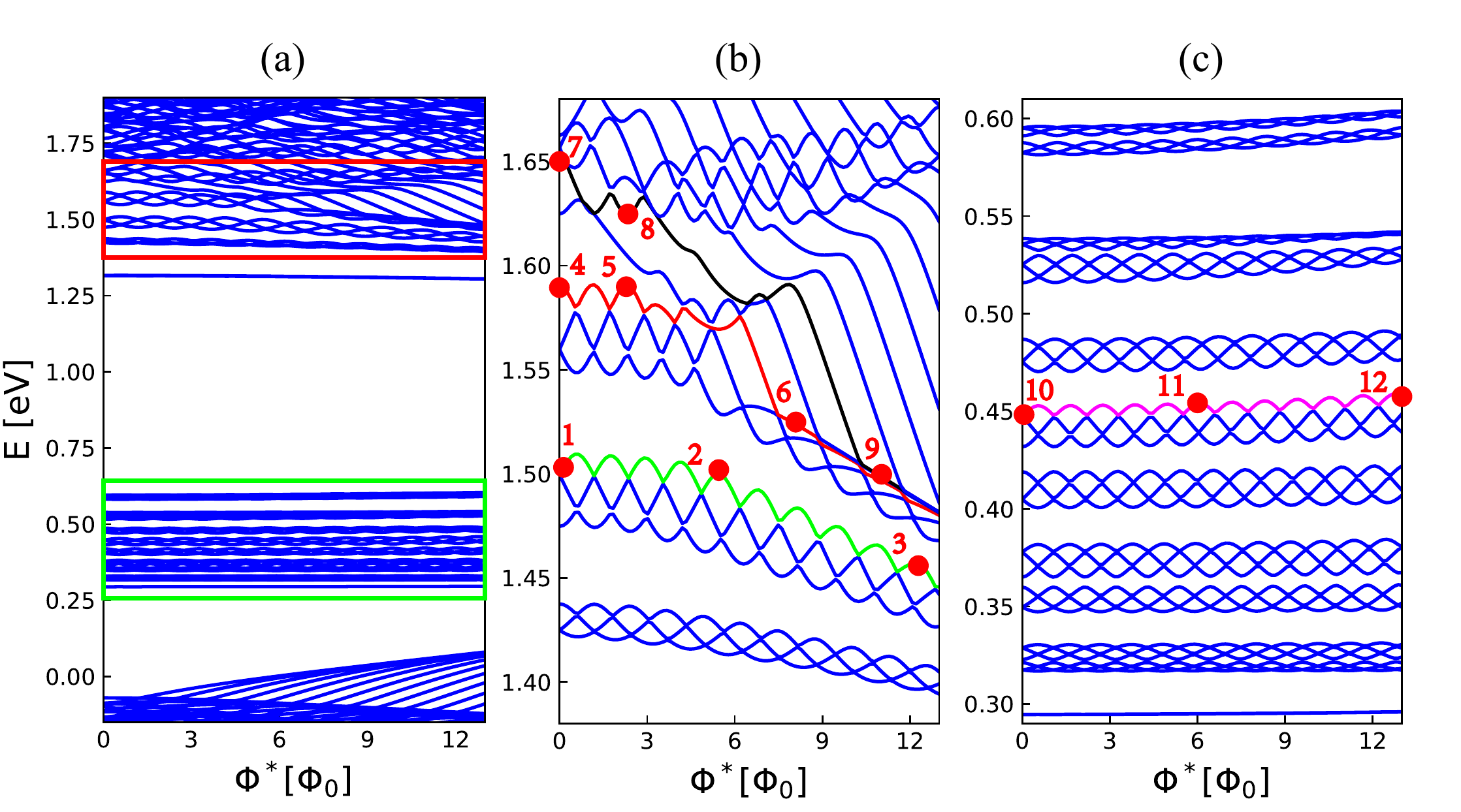}
\caption{ Energy spectrum of HQD, with armchair edge, as a function of the magnetic flux. The regions marked by green and red box are zoomed in (b) and (c), respectively. Green line, red line, black line and pink line in (b) and (c) represent the states $350$, $353$ and $355$ and $326$, respectively.  }\label{fig:fig6}
\end{figure}

Next, we examine the oscillating behavior of HQDs with armchair edges (Fig. 1(c)) which have a higher symmetry than TQDs. The oscillation emerge in two regions in Fig. 6(a). The bottom region, which is marked by the green box, shows perfect oscillation, and the energy levels in top region, which is marked by the red box, bend downward as the magnetic flux increases. In addition, some energy levels only show oscillation in some magnetic region. The oscillating behavior of the energy levels in the red box are different from the energy levels in the green box. In order to explain these behaviors, we plot the corresponding probability densities of some states, which are marked by red dots and numbers on color lines, in Fig. 7. First, we examine the probability densities of the red dots $1$, $4$ and $7$. It is easy to find that $1$ and $4$ are ring states and oscillations of the energy with magnetic field emerge. However, dot $7$ represents a bulk state, and the energy of the bulk state bends downward with increasing magnetic field. The energy spectrum of ring states also bend downward due to the anticrossings of energy levels. The energy of states $350$ and $353$ oscillate with flux and are ring states. However, these ring states are affected by the magnetic flux. The probability densities begin to diffuse to the inner part of HQD from the edges, as is shown by the probability density corresponding to the red dot $3$. The probability density of dot $6$ distribute around the center of HQD, and the ring states become a bulk state. Therefore, the oscillation disappears due to the collapse of ring states. For dot $8$, the state $355$ displayes an oscillating behavior, and the corresponding state is a ring state. The oscillation disappears at dot $9$, because the ring state evolves into a bulk state. In this region, the bulk states and the ring states can be converted into each other by varying the magnetic field. However, the corresponding probability densities of dots $10$, $11$ and $12$ do not change much. The ring states in this region are much more stable under perpendicular magnetic field. Therefore, perfect oscillations emerge.
\begin{figure}
\centering
\includegraphics[width=0.45\textwidth]{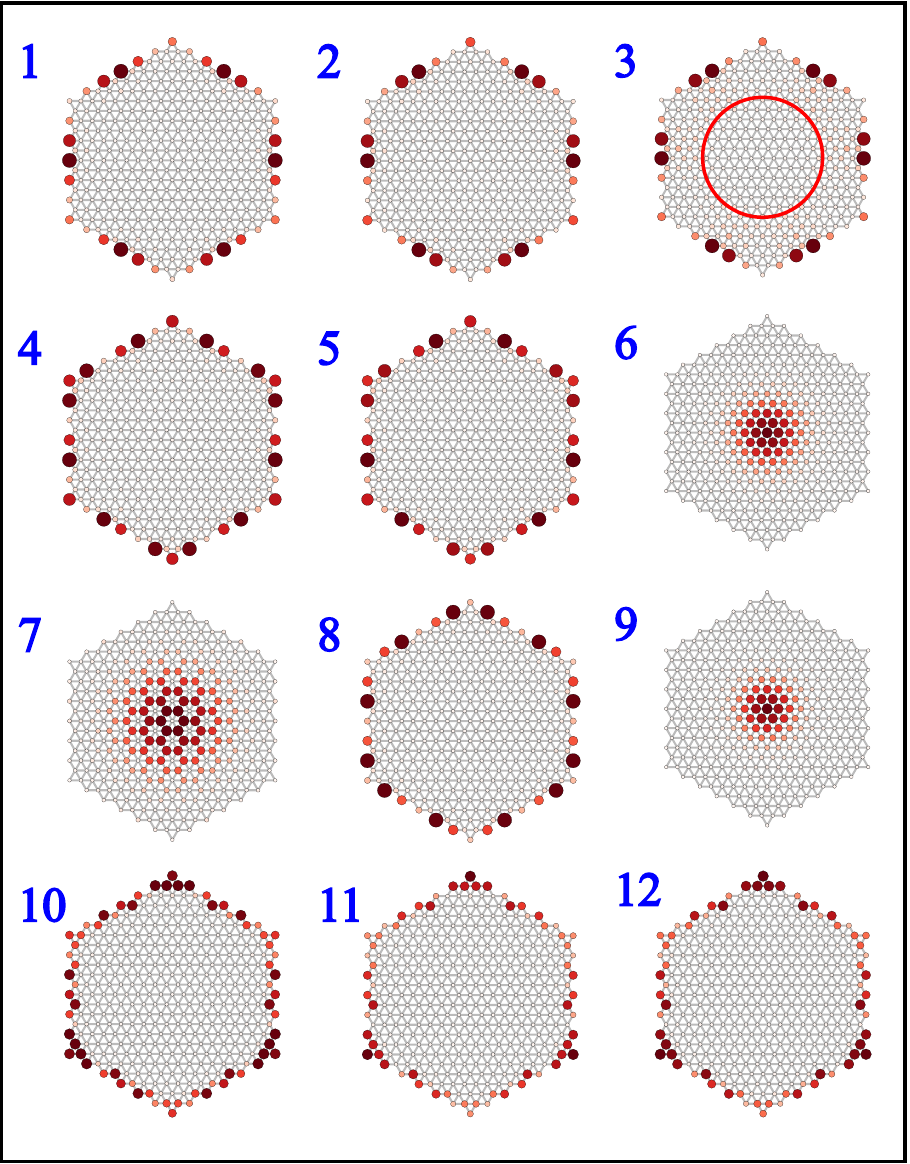}
\caption{ The corresponding probability densities of the states marked by red dots and numbers in Fig. 6.}\label{fig:fig7}
\end{figure}
\begin{figure}
\centering
\includegraphics[width=0.75\textwidth]{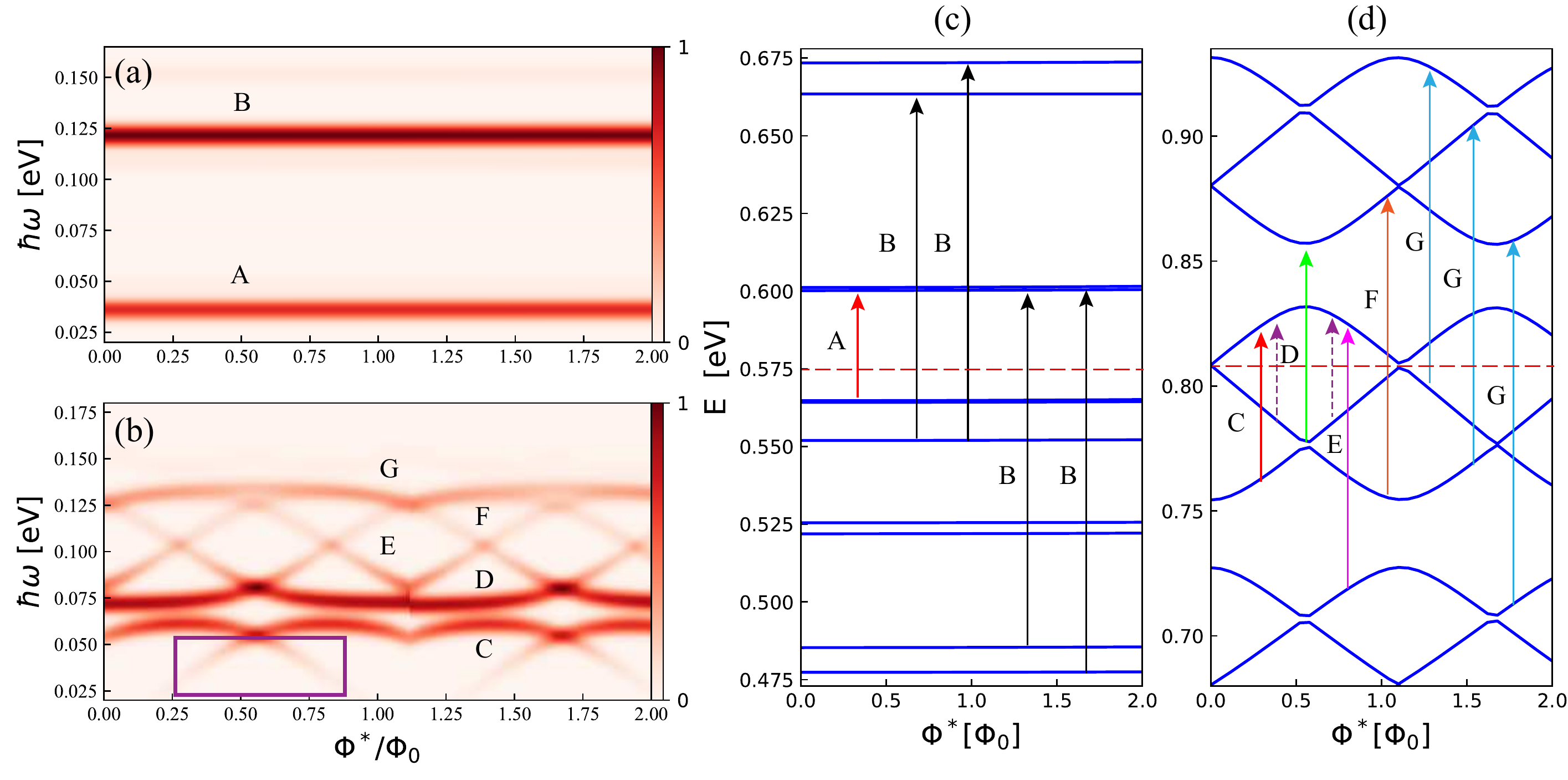}
\caption{ The contour plot of the normalized magneto-optical absorption spectrum of (a) the SQD and (b) the TQD as in, respectively, Figs. 1(a) and (b) with the broadening factor $0.005eV$. The different absorption lines are marked by $A$, $B$$\cdots$ and $G$. The corresponding transitions are marked in (c) for SQD and in (d) for TQD. The red dashed lines represent the Fermi energy, where $E_F=0.575 eV$ for SQD and $E_F=0.809 eV$ for TQD was taken.  }\label{fig:fig8}
\end{figure}

Finally, we examine the optical absorption spectrum of SQD and TQD in the presence of a magnetic field. Here, we tune the Fermi energy inside the gap by a gate voltage and focus only on the optical transition between the edge states. The magnetic field dependence of the optical absorption spectrum is nearly flat in SQD. The optical absorption spectrum in TQD oscillate with the magnetic flux, i.e., it shows AB like oscillations. However, the period of the optical absorption decrease to $1/2$ compared to the period of energy levels. The reason is that the energy level is axisymmetric in each period. There are two strong absorption lines in the optical absorption spectrum of SQD. These two strong absorption lines are caused by transitions between different levels, which are marked in Fig. 8(c). There are more absorption lines in the optical spectrum of TQD, which are caused by transitions between energy levels marked in Fig. 8(d). The absorption tail (purple box) are caused by the transition between the two closest energy levels to the Fermi level, which are marked by dashed purple lines in Fig. 8(d).

\section{The QDs with spin-orbit coupling}
In contrast to some other 2D materials such as graphene and phosphorene, the spin-orbit coupling(SOC) in MoS$_2$ can be large due to the heavy transition-metal Mo atom. For simplicity, here we introduce in the tight-binding model the SOC term by considering only the on-site contribution from the Mo atom, and the strength of SOC is characterized by the parameter $\lambda=0.073 eV$\cite{Yao}. In Fig. 9(a), we found that each energy level is now split into two levels due to the presence of SOC, thereby leading to two sets of energy levels which exhibit Aharonov-Bohm (AB) like oscillations. However, the main oscillating features of the energy levels (e.g., the oscillation period and phase) are not changed significantly by SOC. This is because these energy levels correspond to ring-like edge states which are almost unaffected by SOC, as can be verified by comparing probability densities of such edge states in the absence and presence of SOC.
\begin{figure}
\centering
\includegraphics[width=0.45\textwidth]{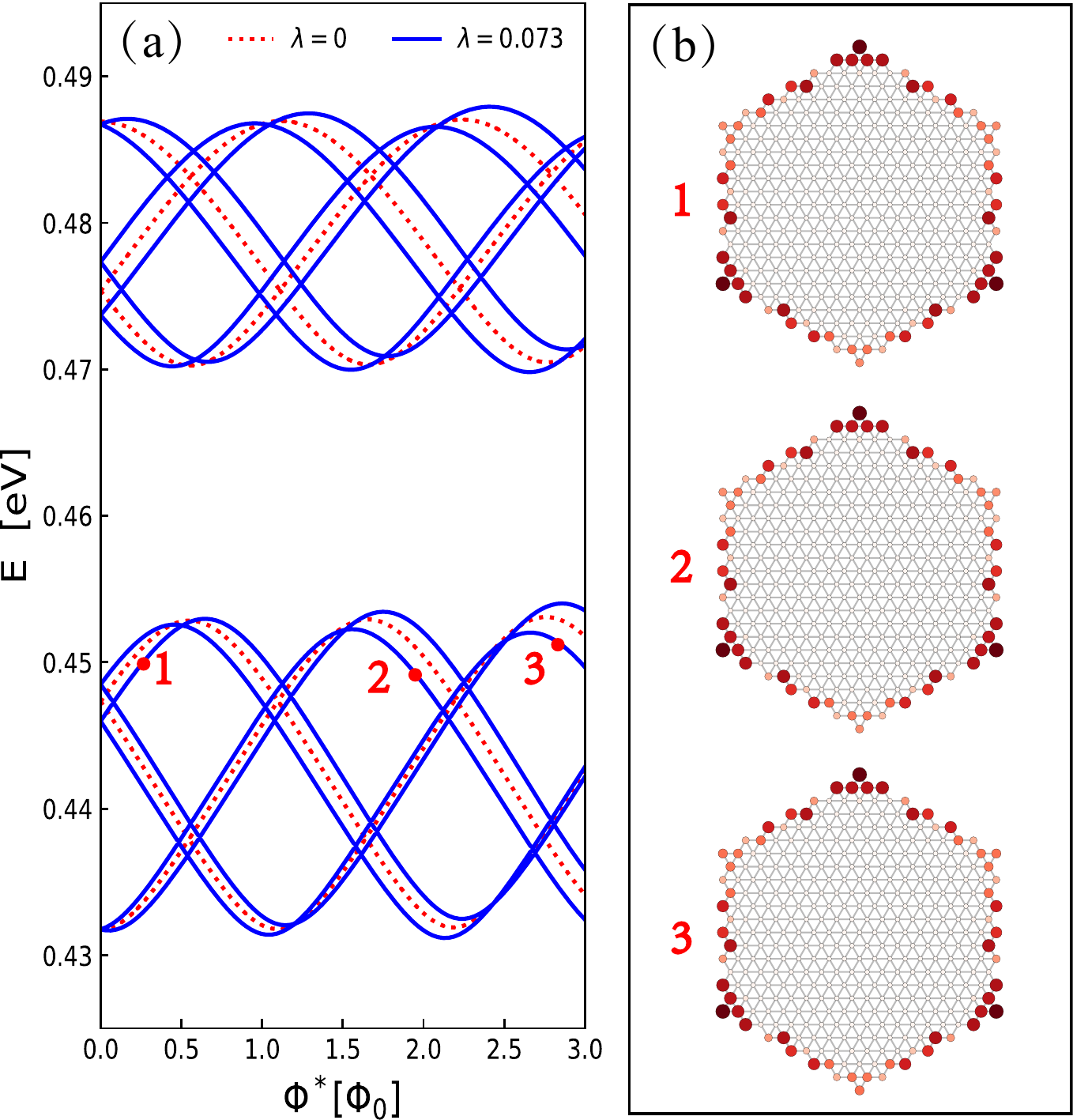}
\caption{ (a) The energy levels without SOC and with SOC, (b) The corresponding probability densities of the states marked by red dots and numbers in (a). The unit of $\lambda$ is $eV$.}\label{fig:fig9}
\end{figure}

\section{summary}
We have investigated the electronic properties of square, triangular and hexagonal quantum dots with armchair and zigzag edges and the influence of a perpendicular magnetic field on the energy spectrum. The energy spectra and wave function of monolayer MoS$_2$ QDs are obtained by solving the tight-binding model. Edge states are found in the gap as in graphene and phosphorene QDs. However, the energy levels of gap states in MoS$_2$ QDs exhibit large differences from graphene and phosphorene QDs, and exhibit oscillating behavior as function of applied field. The origin of those oscillations is that they are ring states that encompass a certain flux. In MoS$_2$ TQDs and HQDs, most gap states are ring states, whose probability is distributed over the whole edge and therefore enclose a well defined flux. The energy levels of ring states oscillate with the magnetic flux, and the period and magnitude of oscillation decrease with the size of the QD. For MoS$_2$ HQDs, there are some ring states that mix with the bulk states in the region close to the conduction band. These bulk and ring states in this region can be converted into each other by the action of the magnetic field. The oscillating dependence on magnetic field disappears as the ring states collapse into a bulk state. Therefore, the oscillations of energy levels in this region are no longer perfect periodic  due to the coupling between ring and bulk states. However, the ring states are robust in the presence of SOC. The oscillation of energy levels also leads to AB like oscillations in the optical absorption spectrum. However, the edge states in MoS$_2$ SQDs are localized states, which are almost unaffected by the magnetic field. In addition, armchair and zigzag edges affect the edge states differently due to the different bonding structure.

\begin{acknowledgments}
Q. Chen acknowledges financial support from the China Scholarship Council (CSC). This work was also supported by Hunan Provincial Natural Science Foundation of
China (No. 2015JJ2040) and by the Scientific Research Fund of Hunan Provincial Education Department (No. 15A042). Additional support from the FLAG-ERA TRANS-2D-TMD. is acknowledged.
\end{acknowledgments}

\end{document}